\documentclass[twocolumn,aps,pra,showpacs,showkeys,preprintnumbers,amsmath,amssymb]{revtex4}
\usepackage{graphicx}
\usepackage{dcolumn}
\usepackage{bm}
\usepackage{color}
\usepackage[usenames,dvipsnames]{xcolor}
\usepackage{caption,graphicx}
\begin{document}
\title{Acoustic Eaton Lens Array and Its Fluid Application}
\author{Sang-Hoon  \surname{Kim}$^{a}$} \email{shkim@mmu.ac.kr}
\author{Sy  \surname{Pham-Van}$^{a}$}
\author{Mukunda P.  \surname{Das}$^{b}$}
\affiliation{
$^a$Division of Marine Engineering, Mokpo National Maritime University,
Mokpo 58628, R. O. Korea
\\
$^b$Department of Theoretical Physics, RSPE, The Australian National University,  Canberra, ACT 0200, Australia
}
\date{\today}
\begin{abstract}
 A principle of an acoustic Eaton Lens array and its application as a removable tsunami wall is proposed theoretically.
The lenses are made of expandable rubber pillars or balloons and create a stop-band by the rotating the incoming tsunami wave and reduce the pressure by canceling each other.
The diameter of each lens is larger than the wavelength of the tsunami near the coast, that is, order of a kilometer.
 The impedance matching on the border of the lenses results in little reflection.
Before a tsunami, the balloons are buried underground in shallow water near the coast in folded or rounded form.
 Upon sounding of the tsunami alarm, water and air are pumped into the pillars, which expand and erect the wall above the sea level within a few hours.
After the tsunami, the water and air are released from the pillars, which are then buried underground for reuse.
 Electricity is used to power the entire process.
 A numerical simulation with a linear tsunami model was carried out.
\end{abstract}
\maketitle
\section{introduction}

A tsunami is a series of one or more massive water waves heading towards the coast. Tsunamis commonly accompany an earthquake that occurs in the ocean and usually cause damage far greater than that caused by the earthquake itself when both impact the land. Various causes of tsunami include volcanic eruptions, asteroid impacts, seabed topography faults, and nuclear explosions. Sometimes a tsunami is classified as a soliton on the surface of a deep sea with a long duration that can be understood by the solution of a nonlinear equation\cite{wang,chabchoub}. Tsunamis are different from the rough waves, which are observed in the open sea, while the former are observed only at the coast. Generally tsunamis are known to be seismic sea waves caused by the displacement of large volume of water appeared on the coast.

The velocity of the tsunami propagation depends on the sea depth $h$ and is estimated
 using the dispersion relation of wave trains on the water surface
  $\omega=\sqrt{gk \tanh(kh)} \simeq k\sqrt{gh}$.
 The value of $kh$ is 1 or less, where $k$ is the wave number\cite{peli,watts}.
 The group velocity is, then, $v=d\omega/dk \simeq \sqrt{gh}$,
where $g$ denotes the gravitational acceleration.
 Because the mean depth of the Pacific Ocean is about 4 km, the speed of a deep sea tsunami through the ocean is approximately 700 km/h.
 Although the speed varies depending on the conditions, it is usually estimated to be similar to that of a passenger aircraft.

 As the tsunami enters the shallow waters of the coastal areas,
the velocity of the tsunami slows and its wavelength is shortened by one or two kilometers, but the amplitude grows enormously.
 The wave front also becomes parallel to the coastline.
Water waves are made up of a combination of longitudinal and transverse waves.
 When a water wave comes into shore, the vertical height above sea level to which it rises is called the run-up.
 The high run-up of multiple waves together with the long inundation distance of the tsunami
 makes it one of the most devastating ocean disasters.
  Large events can generate wave heights of tens of meters.
 The majority of the damage from a tsunami comes from the smashing force and destructive power
 due to the high pressure of a large volume of water\cite{charus,ahmad,triat}.
It is extremely difficult for a hard wall to withstand the huge momentum of a tsunami.
A concrete wall does not guarantee the safety of the coastal infrastructures behind the wall.
The combination of smashing forces and high pressure can impose serious damage to the stability of a hard wall.
Though tsunamis have been studied quite extensively,
 protection methods other than the concrete wall method are very rare.
For protection from earthquakes, some methods incorporating metamaterials have been suggested\cite{kim3,kim4}.

Recently a couple of theoretical proposals for a tsunami wall were introduced.
One is an acoustic cloaking method\cite{farhat}.
 This method utilizes an acoustic application of the invisibility cloak that has been described for electromagnetic waves and is applicable
because electromagnetic and surface water waves are analogous.
The other proposed solution is a sea tube\cite{hu}.
 The sea tube is a two-dimensional periodic concrete tube array placed in the shallow water near the coast.
The tube is an empty pillar or cylinder that has vertical slits on the surface
 that allows the water to move in and out of the tube to reduce the water pressure by resonance.

 In this paper  we present a simple theoretical proposal of a removable rubber tsunami wall
based on gradient index (GRIN) lens technology that originated in optics.
It is a {\it tsunami-against-tsunami} method with an array of flexible acoustic Eaton lenses (AEL) used to produce wave rotation.
The principle is to cancel out the incoming wave into a rotating wave.
 Impedance matching results in little reflectance of the wave.
Before the tsunami arrives, there is nothing above sea level.
 Upon sounding of the tsunami alarm, the tsunami walls are erected above the sea level within a few hours using electric power.
The wall reduces the water pressure of the tsunami.
After the tsunami, the wall is lowered down to underground within a few hours using electric power.

\section{acoustic Eaton lens and Tsunami modeling}

An Eaton lens is a GRIN lens in which the refractive index (RI) is given as a function of the radius.
It is an analogue of Kepler's scattering problem, which is derived from the relation
between light trajectory and its RI at any impact parameter\cite{tyc,ma,hendi}.
It has a singularity in which the RI goes to infinity at the center of the lens and the speed of light is reduced to zero at this point.
Then, in principle, the lens can change the wave trajectory into any direction.
 Impedance matching prevents reflection on the border.
Compared with the classical curved lenses, GRIN lenses may have flatter shapes and are free from geometric aberrations.

The RI of the Eaton lens at arbitrary refraction angles is given by Eq.~(\ref{100})\cite{kim}.
\begin{equation}
 n^{\pi/\theta} = \frac{R}{nr}+\sqrt{\left( \frac{R}{n r}\right)^2-1},
\label{100}
\end{equation}
where $\theta$ is any refraction angle as a radian,
$R$ is the radius of the lens, and $r$ the distance to the center.
Note that  $n=1$ for $r \ge R$.
  The most well-known Eaton lens is the simplest form of  $\theta=\pi$
   in Eq.~(\ref{100})
\begin{equation}
 n = \sqrt{\frac{2R}{r}-1}. 
\label{102}
\end{equation}

The 180-degree Eaton lens in Eq.~(\ref{102}) is a kind of mirror in which the incident wave returns towards the original direction without being upside down\cite{ma}.
The rotating property of the Eaton lens appears regardless of its oscillating direction to that of the energy transfer.
This Eaton lens has been realized in optics using the variable electrical permittivity method\cite{zent}.
 The acoustic analogy of the Eaton lens is the key to the Tsunami wall
because its lenses bend acoustic waves including seismic and water waves in almost any direction\cite{guenneau}.
If we send arbitrary waves towards an Eaton lens, the trajectories of the waves
overlap as a result of traveling in different directions and their intensity is reduced.
This is the case for any mechanical waves whether they are longitudinal or transverse.

\begin{figure}
\centering
\resizebox{!}{0.25\textheight}{\includegraphics{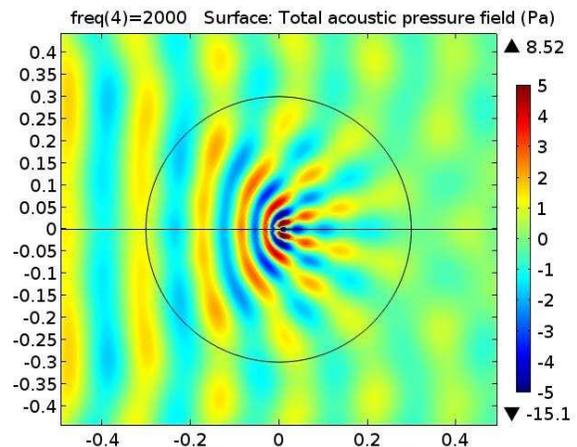}}
\caption{
Acoustic pressure by the analytic Eaton lens. f=2,000Hz and the unit is m.
\label{theory}}
\end{figure}

\begin{figure}
\centering
\resizebox{!}{0.27\textheight}{\includegraphics{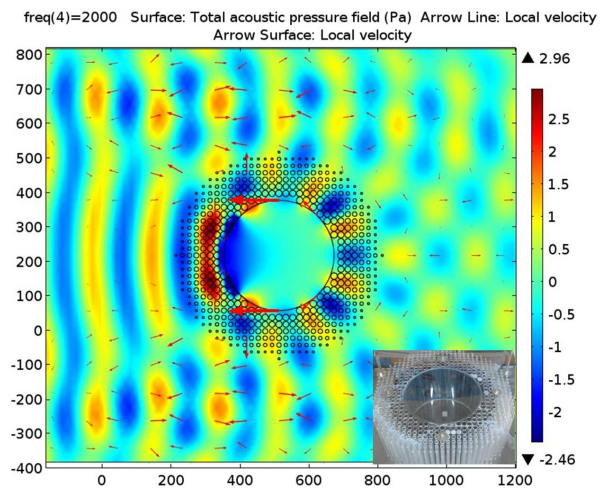}}
\caption{Acoustic pressure fields by the designed acoustic Eton lens and a real sample.
 f=2,000Hz and the unit is mm.
\label{design}}
\end{figure}

Tsunami is a mechanical wave that propagates along the interface between water and air
 and, then, nonlinear wave in general.
 The basic hydrodynamic model of tsunami restored by gravity is obtained by
the conservation of momentum of incompressible fluids.
It is the Navier-Stokes equation\cite{farhat,hu,comsol}:
\begin{equation}
\rho \left( \frac{\partial }{\partial t}+{\bf u} \cdot \nabla \right){\bf u}
-\mu \nabla^2 {\bf u}=-\nabla p + \rho {\bf g},
\label{1}
\end{equation}
where $\rho$ is the density, $\mu$ is the viscosity, and $p$ is the pressure.
${\bf u}$ and ${\bf g}$ are the velocity and gravitational fields.
Fluctuating two-dimensional linear approximation has been used for tsunami except for resonances\cite{peli,watts}.
We restrict our model to an inviscid linear wave of infinite extent with very thin depth.
Assuming a time-dependent harmonic wave, the pressure varies with time as
$p({\bf r},t) =p({\bf r})e^{i \omega t} $, where $\omega$ is the angular frequency.
For the harmonic water waves, the elevation of water surface $\eta$
is given by
$\eta({\bf r},t) = Re \left\{ -\frac{i \omega}{g} \phi({\bf r}) e^{- i \omega t} \right\}$,
where $\phi$ is the potential which satisfies the two-dimensional Helmholtz equation on the free surface.
\begin{equation}
(\nabla^2 + k^2)\phi = 0,
\label{3}
\end{equation}
with $k$ as the wave number.
Commercial simulator of COMSOL MULTIPHYSICS 5.1 was used for the numerical simulation\cite{comsol}.

The numerical simulation of the wave cancelation by the analytical acoustic Eaton lens is shown in FIG.~\ref{theory}\cite{youtube1}.
  The lens bends the wave of shorter wavelength compared with its diameter effectively.
Therefore, the diameter of the lens should be larger than the wavelength of the tsunami near the coast.
The material of the Eaton lens was chosen acrylic,
but any material with large impedance compared with that of the background produces similar result in the numerical simulation.

\begin{figure}
\centering
\resizebox{!}{0.26\textheight}{\includegraphics{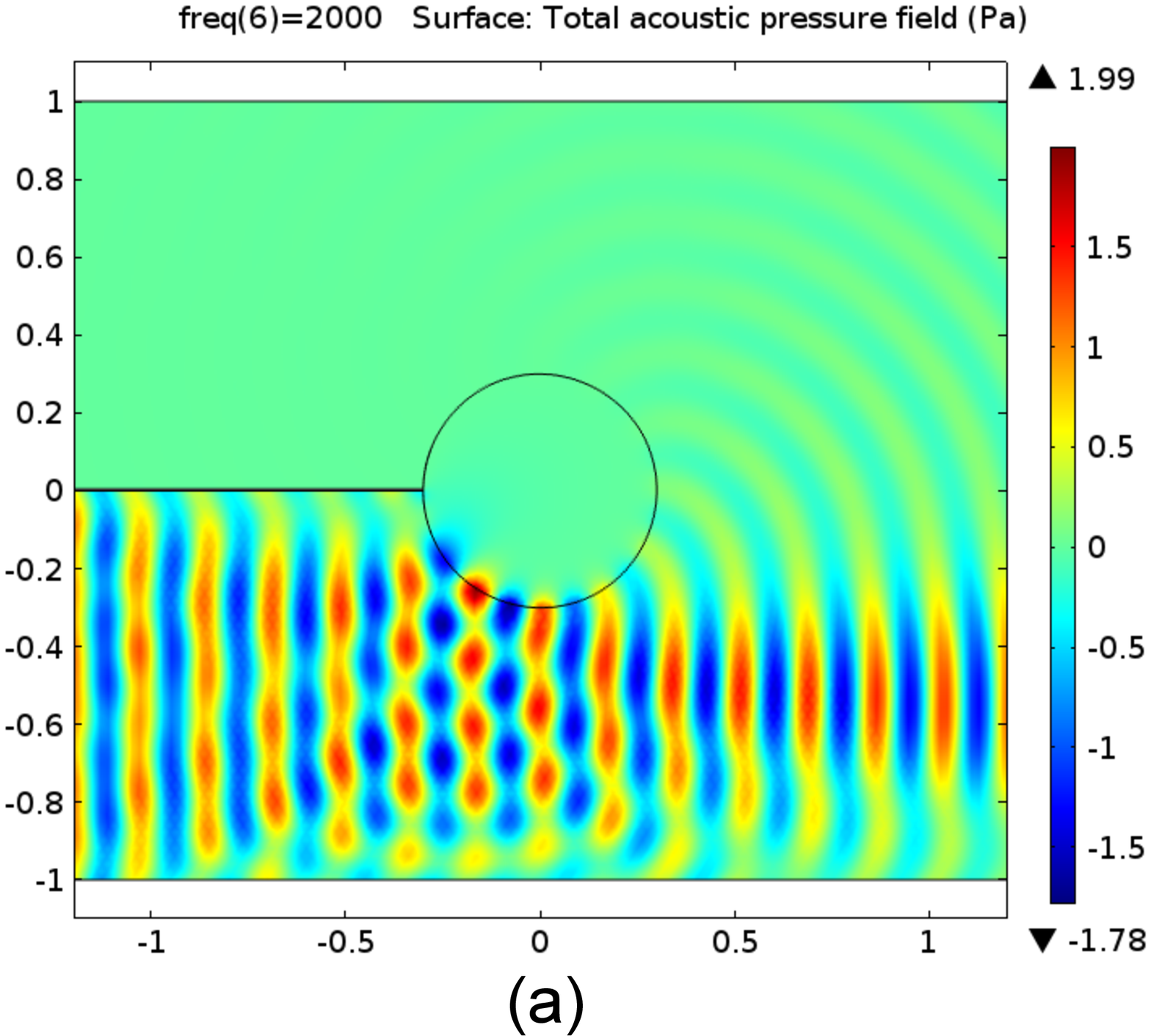}}
\resizebox{!}{0.26\textheight}{\includegraphics{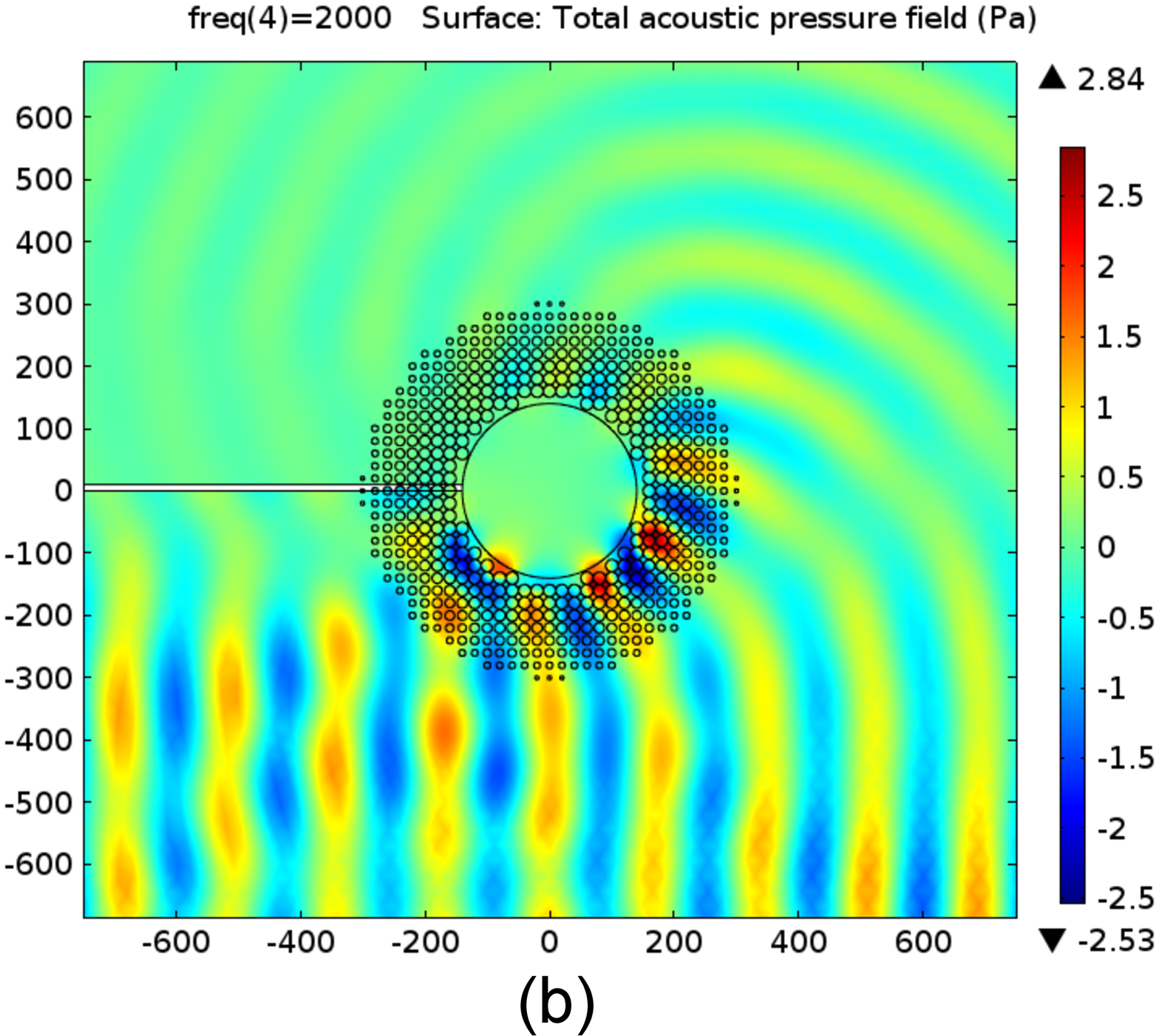}}
\caption{Demonstration of the rotating ability and reflection. (a) Concrete cylinder. The unit is m.
(b) Acoustic Eaton lens. f=2,000Hz and the unit is mm.
\label{rotation}}
\end{figure}

 The wave equation of the acoustic counterpart is governed by the density and bulk modulus or the inverse compressibility of the medium.
The background speed of a sound wave is $v_o=\sqrt{B/\rho_o}$,
where $B$ and $\rho_o$ are the background bulk modulus and density outside the lens, respectively.
The modulus is assumed to be constant.
Therefore, the variable density method is applied as\cite{kim5}
\begin{equation}
n(r) = \frac{v_o}{v(r)} = \sqrt{\frac{\rho(r)}{\rho_o}}.
\label{4}
\end{equation}
The refractive index in Eq.~(\ref{102}) is digitalized and matched with in Eq.~(\ref{4}) as
\begin{equation}
 \frac{2N}{i-0.5}-1 = \frac{\rho_i}{\rho_o},
\label{20}
\end{equation}
where i=1,... N.

Numerical simulations of the designed acoustic Eaton lens is shown in FIG.~\ref{design}.
A model of an  AEL was designed and made of acrylic
as the wall shown in FIG.~\ref{design} inserted. It shows the overall shape of the lens.
N=15 was chosen in Eq.~(\ref{20}).  R=30 cm and the distance between pillar is 2 cm.
The diameter of the center circle is 32 cm or approximately half that of the lens.
Beyond the center, the circular columns of the diameter between 19 mm and 6 mm are radially located.
The refractive index of the Eaton lens has a singularity at the center or $r=0$ in Eq.~\ref{102}.
It is a cut at some point in design and the part becomes a large single column in the center.
Our choice of the cut was the half of the radius, but it makes the impedance matching incomplete.
There are 488 small pillars around the large one and the number of the pillars is adjustable.
The more the pillars, the better for the rotation and the impedance matching, but more expensive to build.

\begin{figure}
\centering
\resizebox{!}{0.22\textheight}{\includegraphics{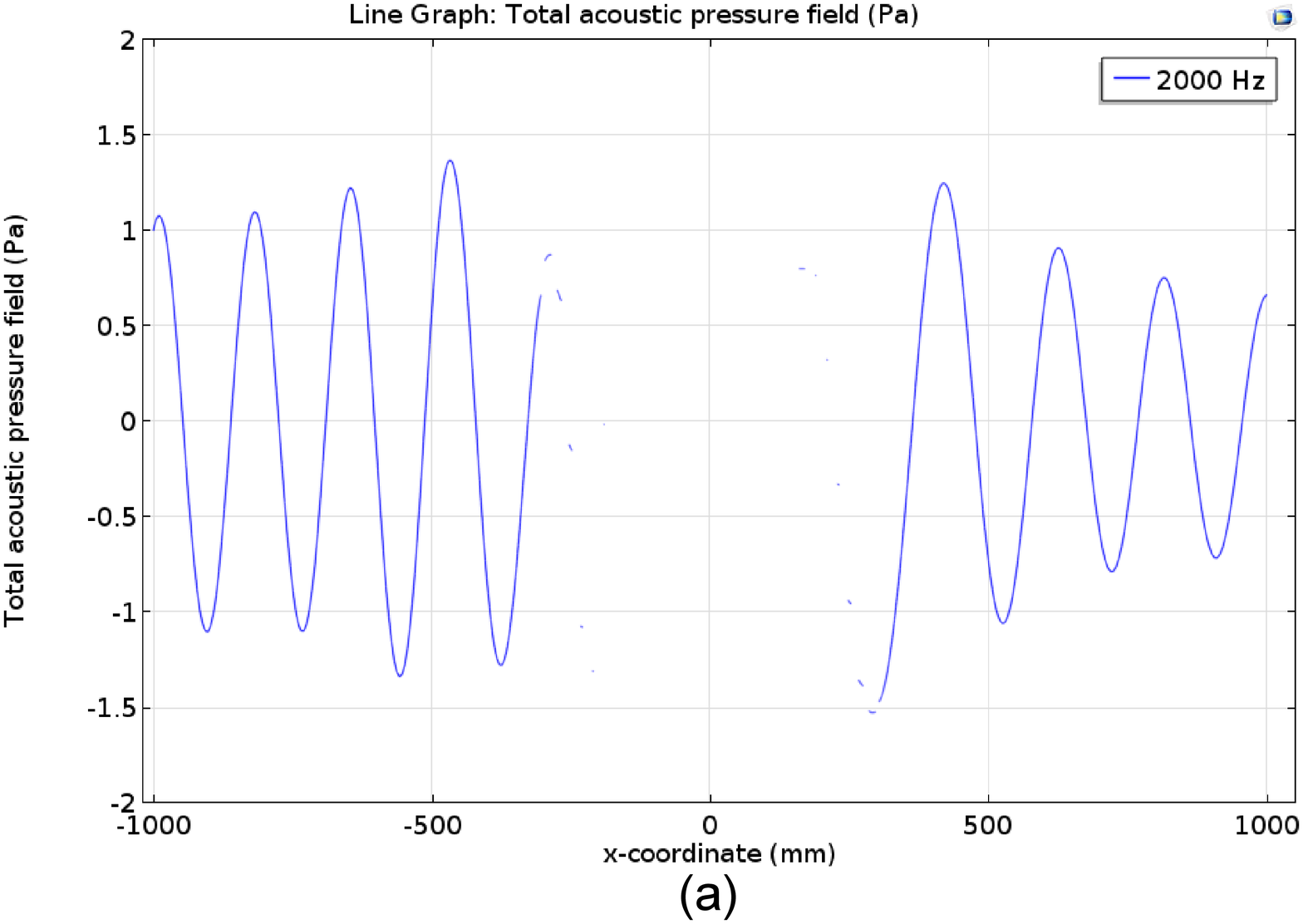}}
\resizebox{!}{0.22\textheight}{\includegraphics{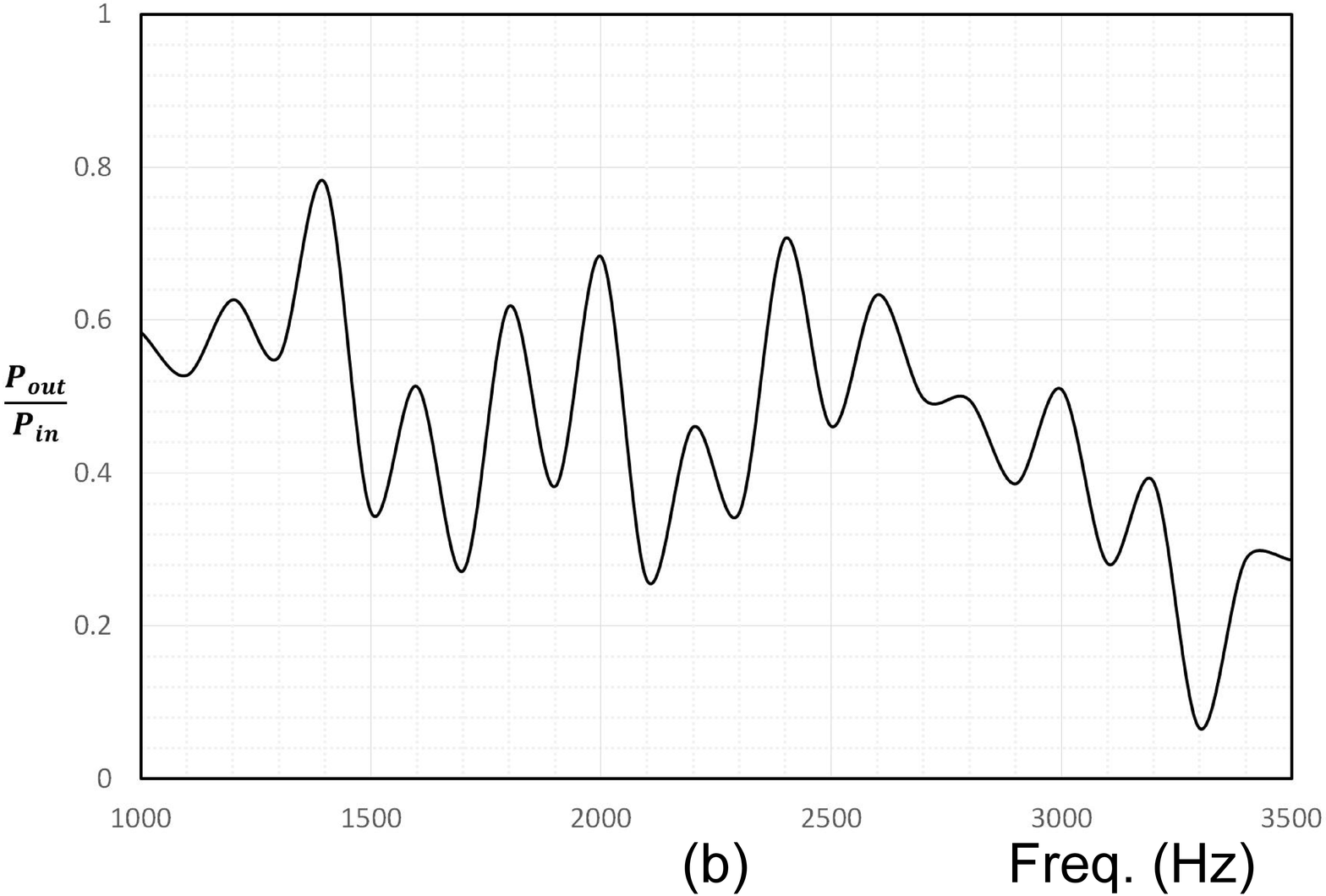}}
\caption{(a) Acoustic pressure just before and after an Eaton lens. f=2,000Hz and the unit is mm.
(b) Frequency dependent pressure ratios between input and output waves.
\label{ratio}}
\end{figure}

The rotating ability and reflectance of the designed AEL must be confirmed and compared with those of a concrete cylinder.
 Such a simulation is plotted in FIG.~\ref{rotation}.
 A plane wave was applied in FIG.~\ref{rotation} for convenience but the lens works for any types of waves regardless of linear or nonlinear wave\cite{youtube2}.
It is easier to induce rotation in waves with a short wavelength.
 The AEL in  FIG.~\ref{rotation}(b) produces strong rotation and little reflection, unlike the concrete cylinder in  FIG.~\ref{rotation}(a).

The acoustic pressure at the center line of the lens is plotted in FIG.~\ref{ratio}(a).
There is no acoustic pressure between -320 $\sim$ 320 mm because it is the center region of large impedance difference and that wave cannot pass through.
It assumed to be very small compared with the outside region.
The pressure ratios between input and output waves away from the scatter but same distances are plotted at several applied frequencies in FIG.~\ref{ratio}(b).
The output pressure becomes one half that of the input pressure after passing through just one AEL (average).
Therefore, a double array of Eaton walls will reduce the output pressure about one quarter.
The AEL is not very effective in reducing the acoustic pressure for a wavelength larger than the diameter of the lens.

\section{Tsunami wall engineering}

\begin{figure}
\centering
\resizebox{!}{0.15\textheight}{\includegraphics{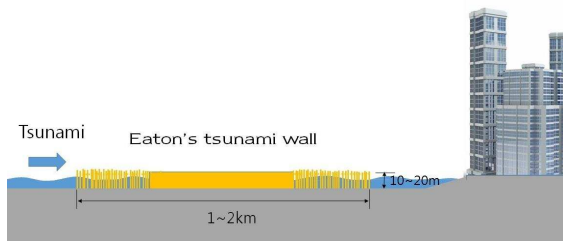}}
\caption{A model view of the tsunami wall with one acoustic Eaton lens.
\label{view}}
\end{figure}
\begin{figure}
\centering
\resizebox{!}{0.26\textheight}{\includegraphics{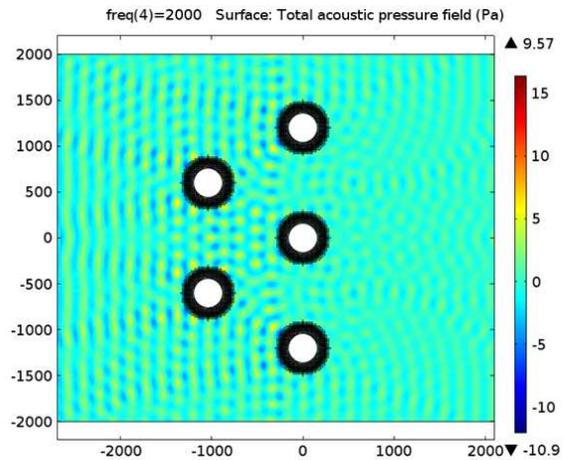}}
\caption{2+3 two-layer Eaton wall reducing the pressure into one quarter. f=2,000Hz.
\label{wall}}
\end{figure}

Our tsunami wall comprises AEL arrays that are made from expandable rubber pillars or balloons.
The pillars are buried underground in shallow water near the coast in a folded or rolled form.
Each lens is connected to underground rubber pipes.
Each pillar of the lens behaves like a rubber balloon.
The large center region is divided into many rubber pillars and held together by ropes.
There is an illustration to build an Eaton wall in FIG.~\ref{view}.
Make two concentric circles of diameter 900 m and 1,800 m at the boundary between sea and land.
Build rubber pillars tightly following the perimeter of the small circle not to pass through the tsunami wave as possible as or the inside of the small circle is composed of many pillars and held together by ropes.
The height of the pillars is 10$\sim$20 m.
Make a two-dimensional square lattice of lattice constant 60 m between the two circles.
Each lattice point has one rubber pillar of diameter from 57 m to 18 m in the radial direction following Eq.~(\ref{20}).
Then, one Eaton lens of about 500 pillars is built.

The wave-against-wave method is converted into a tsunami-against-tsunami method.
A sample array of the lenses with equal diameters is shown in FIG.~\ref{wall}.
The distance between boundaries of the lenses must be smaller than the wavelength of the tsunami or order of km.
 Shorter wavelength tsunamis are less dangerous, so we allow them to pass between the lenses.
Before the tsunami alarm, the walls are buried underground in shallow water near the coast in a folded or rolled form.
Each pillar is connected to underground rubber pipes for pumping water and air.
 When the tsunami alarm is sounded, water and air are pumped into the pillars and they expand and erect above the sea level within a few hours.
The lenses puff up like rubber balloons.
The array of lenses acts as a tsunami wall to reduce the water pressure.
After the tsunami, the water and air are released from the pillars
 and they return underground for reuse.
 The entire process is operated by electric control.

\section{Summary and Discussion}

A theoretical removable tsunami wall that is based on an optical Eaton lens is proposed to control tsunami waves.
The principle is to transform the incoming wave into a rotating wave.
A model acoustic Eaton lens was designed using the variable density method.
We used 180 degree Eaton lenses, but Eaton lenses greater than 90 degrees show a similar pressure-reducing effect.
The wall is an array of acoustic Eaton lenses comprised of many hundreds of flexible rubber pillars, and works as a stop-band for tsunami.
The impedance matching is used to reduce the pressure of the tsunami without reflecting the wave.
 The rubber-made wall is reusable and easy to erect and disassemble.
 The costs of construction and maintenance are very low compared with the conventional methods.

Tsunami is an unexpected disaster in nature and there have been numerous studies for that. Nonetheless, we humans have been almost defenseless except early warnings or building a huge barrier. The momentum of tsunami is too huge to control and it will be a wild suggestion to defend tsunami with acoustic metamaterial method originated from geometrical optics.
This research is just the beginning for that and still far away to apply in the real world of civil Engineering.
We have a modest plan to test the wall with miniatures in a pool sometime.
We hope that this research would be a first step for scientists and engineers to the challenge of tsunami.

\acknowledgments
S.-H. Kim thanks to Prof. J. S$\acute{a}$nchez-Dehesa for useful discussions.


\begin{thebibliography}{0}

\bibitem{wang} X. Wang and P. L.-F. Liu,
``Numerical Simulation of the 2004 Indian Ocean Tsunamis - Coastal Effects,"
J. of Earthquake and Tsunami, {\textbf 1}, 273-297 (2007).

\bibitem{chabchoub} A. Chabchoub, O. Kimmoun, H. Branger, N. Hoffmann, D. Proment, M. Onorato, and N. Akhmediev,
``Experimental Observation of Dark Solitons on the Surface of Water," 
Phys. Rev. Lett.,
{\textbf 110}, 124101(1-5) (2013).
\bibitem{peli} E. Pilinovsky, T. Talipova, A. Kurkin, and C. Kharif,
``Nonlinear mechanism of tsunami wave generation by atmospheric disturbances,"
Natural Hazards and Earth System Sciences, {\textbf 1}, 243--250 (2001).

\bibitem{watts} P. Watts, S. T. Grilli, J. T. Kirby, G. J. Fryer, and D. R. Trappin,
``Landslide tsunami case studies using a Boussinesq model and a fully nonlinear tsunami generation model,"
Natural Hazards and Earth System Sciences, {\textbf 3}, 391--402 (2003).

  \bibitem{triat} R. Triatmadja and A. Nurhasanah,
``Tsunami Force on Buildings with Openings and Protection,"
J. of Earthquake and Tsunami, {\textbf 6}, 1250024(1--17) (2012).

\bibitem{charus} N. Charusrojthanadech, Y. Yamamoto, and V. T. Ca,
``Trial of an Evaluation Method of the Building Damage by Tsunami,"
J. of Earthquake and Tsunami, {\textbf 6}, 1250014(1--31) (2012).

\bibitem{ahmad} S. M. Ahmad and D. Choudhury,
``Stability of Wavefront Retaining Wall Subjected to Pseudo-Dynamic Earthquake Forces and Tsunami,"
J. of Earthquake and Tsunami, {\textbf 2}, 107--131 (2008).

\bibitem{kim3} S.-H. Kim and M. P. Das,
``Seismic Waveguide of Metamaterials,"
Mod. Phys. Lett. B, {\textbf 26}, 1250105(1--8) (2012).

\bibitem{kim4} S.-H. Kim and M. P. Das,
``Artificial Seismic Shadow Zone of Acoustic Metamaterials,"
Mod. Phys. Lett. B, {\textbf 27}, 1350140(1--9) (2013).

\bibitem{farhat} M. Farhat, S. Enoch, S. Guenneau, and A. B. Movchan,
``Broadband Cylindrical Acoustic Cloak for Linear Surface Waves in a Fluid," 
Phys. Rev. Lett.
{\textbf 101}, 134501(1--4) (2008).

\bibitem{hu} X. Hu, C. T. Chan, K.-M. Ho, and J. Zi,
``Negative Effective Gravity in Water Waves by Periodic Resonator Arrays," 
 Phys. Rev. Lett.
 {\textbf 106}, 174501(1--4) (2011).

 \bibitem{tyc} T. Tyc and U. Leonhardt,
``Transmutation of singularities in optical instruments,"
 New J. Phys. \textbf{10}, 115038(1--8) (2008).

\bibitem{ma}  Y. G. Ma,   C. K. Ong, T.  Tyc, and U. Leonhardt,
``An omnidirectional retroreflector based on the transmutation of dielectric singularities,"
Nature, Mat. \textbf{8}, 639--642 (2009).

\bibitem{hendi} A. Hendi, J. Henn, and U. Leonhardt,
``Ambiguities in the Scattering Tomography for Central Potentials," 
Phys. Rev. Lett.
 \textbf{97}, 073902(1--4) (2006).

\bibitem{kim} S.-H. Kim,
``Retroreflector Approximation of a Generalized Eaton lens,"
J. of Modern Optics, {\textbf 59}, 839--842 (2012).

\bibitem{zent}  T. Zentgraf, Y. Liu, M. H. Mikkelson, J. Valentine, and X. Zhang,
 ``Plasmonic Luneburg and Eaton lenses,"
 Nature Nano. {\textbf 6}, 151--155 (2011).

\bibitem{guenneau}  T. M. Chang, G. Dupont, S Enoch, and S. Guenneau,
 ``Enhanced control of light and sound trajectories with three-dimensional gradient index lenses,"
 New J. Phys. {\textbf 14}, 035011(1--29) (2012).

\bibitem{comsol} COMSOL MULTIPHYSICS, {\it Acoustics Module User's Guide}, Ver. 4.3b (comsol,2013) Ch. 3.

\bibitem{youtube1} https://www.youtube.com/watch?v=SWOMDiDAImM.

\bibitem{kim5} S.-H. Kim,
``Cylindrical Acoustic Luneburg Lens,"
 8th Inter. Cong. on Adv. Ele. Mat. in Microwaves and Optics, IEEE Xplore, 364--366 (2014).

\bibitem{youtube2} (a) https://www.youtube.com/watch?v=dnRcWgQeYgw,
 (b) https://www.youtube.com/watch?v=ZcATqTU2s9o.

\end{thebibliography}
\end{document}